# Development of High-Quality α-Ta Film at Room Temperature via Seed Layer Engineering

*Senthil Kumar Karuppannan,[a]\* Naga Manikanta Kommanaboina,[a] Hui Hui Kim,[a] Nelson Lim Chee Beng,[a] Yap Lee Koon Sherry,[a] Yan Guangxu,[a] and Manas Mukherjee[a,b]\**

[a]NQFF, Institute of Materials Research and Engineering (IMRE), Agency for Science, Technology and Research (A*STAR), 2 Fusionopolis Way, Innovis Building #08-03, Singapore 138634, Republic of Singapore.

[b]Centre for Quantum Technologies, National University Singapore, Singapore, 117543, Republic of Singapore.

ABSTRACT: The growth of high-quality superconducting thin film on silicon substrates is essential for quantum computing, and low signal interconnects with industrial compatibility. Recently, the growth of α-Ta (alpha-phase tantalum) thin films has gained attention over conventional superconductors like Nb and Al due to their high-density native oxide ($Ta_2O_5$), which offers excellent chemical resistance, superior dielectric properties, and mechanical robustness. The growth of α-Ta thin films can be achieved through high-temperature/cryogenic growth, ultra-thin seed layers, or thick films (>300 nm). While high-temperature deposition produces high-quality films, it can cause thermal stress, silicide formation at the interface, and defects due to substrate-film mismatch. Room-temperature deposition minimizes these issues, benefiting heat-sensitive substrates and device fabrication. Low-temperature growth using amorphous (defective) seed layers such as TaN and TiN has shown promise for phase stabilization. However, nitrogen gas, used as a source of metallic nitride, can introduce defects and lead to the formation of amorphous seed layers. This study explores using crystalline seed layers to optimize α-Ta thin films, demonstrating improved film quality, including reduced surface roughness, enhanced phase orientation, and higher transition temperatures compared to amorphous seed layers like metal nitrides. These advancements could interest the superconducting materials community for fabricating high-quality quantum devices.

# 1. Introduction

Superconducting thin films deposited on low-dielectric substrates, such as high-resistive silicon, $SrTiO_3$, $LaAlO_3$, MgO, and sapphire, play a crucial role in the fabrication of superconducting material-based quantum devices.[1–4] While various thin films and substrates have been explored for their growth and performance, achieving high-quality thin film deposition on silicon substrates is particularly attractive for industrial integration. Since integrating quantum devices requires substrates compatible with the microelectronics industry, silicon is preferred.[5–7] Although sapphire substrates can support high-quality superconducting device architectures,[8,9] they are incompatible with commercial CMOS technology.

The growth of α-Ta on silicon substrates has recently gained significant importance in the superconducting community due to its ability to form a high-density, stable oxide compared to other superconducting thin films.[10–12] Tantalum exists in alpha (α) and beta (β) phases, with typical room-temperature growth conditions resulting in a mixed phase.[12] The bulk α-Ta phase metal exhibits superconductivity at ~4.5 K, whereas the mixed or β-Ta phase, with its loosely packed structure, becomes superconducting only at much lower (mK) temperatures.[13] Various methods can be employed to obtain α-Ta thin films, including deposition at higher temperatures, the use of seed [12,14,15], increasing film thickness >300 nm[16,17], and deposition at cryogenic temperatures[13].

High-temperature deposition of Ta on a Si substrate can form the α-Ta phase; however, it does not always result in a high-quality film.[18,19] This is because Ta is highly reactive with heated Si, forming metal silicide,[18–22] which can degrade film properties and cause microwave signal leakage—an issue for superconducting and electronic applications. While increasing the film thickness can promote α-Ta growth[16], it is incompatible with device fabrication, as the dry etching process generates excessive heat and surface contamination. For instance, hardened polymer residues become challenging to remove after etching. Therefore, the film thickness should be limited to around 200 nm for superconducting device fabrication. Although α-Ta growth at cryogenic temperatures is an option, it is costly and requires specialized tools. A more practical approach is room-temperature growth using seed layers to achieve α-Ta formation without Ta-silicide formation, ensuring high-quality films. However, commonly used amorphous metal nitride seed layers introduce defects at the interface. Recent studies have shown that heated TiN seed layers perform better than unheated ones due to reduced defects.[23] This highlights the need for alternative seed layers to create a high-quality interface for improved α-Ta film growth. Here, we systematically investigate the effect of seed layers, ranging from amorphous (defective) to crystalline metallic seed layers, on the quality of α-Ta film. XRD, XPS, AFM, and $T_c$ measurements confirm that changing the seed layers from amorphous to crystalline alters the film quality. The following sections provide detailed characterization and explanations.

## 2. Experimental details.

*Materials*

The single-sided polished high-resistive 4-inch silicon (100) wafer, with a resistivity of 20000 Ω.cm and a thickness of 550±25 $\mu$m, was purchased from Latech Scientific Supply Pte. Ltd. The buffered oxide etching solution (7:1 ammonium fluoride ($NH_4F$) to hydrofluoric acid), developer, and other cleanroom solvents were acquired from Merck Ltd. The photoresist S1818 was obtained from MicroChemicals GmbH. The following materials were purchased and utilized in the TIMARIS cluster tool: Ta-ACI alloy (99.95% purity), Ru (99.95% purity, Heraeus), and Ti (99.95% purity, Materion).

*Thin film deposition*

The multi-chamber physical vapor deposition (PVD) TIMARIS cluster tool was used to deposit 5 nm of various seed layers, followed by 100 nm of Ta thin film. First, the silicon wafer was cleaned for 60 seconds in a buffered hydrofluoric acid solution with a 7:1 ratio of $NH_4F$ to hydrofluoric acid, followed by a wash with deionized water to remove contaminants and native oxides from the silicon surface. The wafer was loaded into the PVD system within 20 minutes after cleaning, achieving a process chamber base pressure of $\leq 5 \times 10^{-8}$ Torr. A 100 nm thick layer of Ta metal was deposited using the TIMARIS cluster tool—a vacuum nano deposition system—with a source power of 1.4 kW, a chamber pressure of 0.003335 mbar, a gas flow rate of 400 sccm of argon, a substrate temperature maintained at 25°C, and a deposition rate of 10 nm/min. Before the Ta deposition, in-situ argon plasma was used to clean the substrate in a separate etching chamber. To deposit the 5 nm of various seed layers, the pre-cleaned wafer was transferred to a different deposition chamber without breaking the vacuum, maintaining a central chamber vacuum of $\leq 5 \times 10^{-8}$ Torr throughout the process. The 5 nm of Ti was deposited at a gas flow rate of 300 sccm and a chamber pressure of $2.6 \times 10^{-3}$ mbar. The Ru deposition was conducted with a source power of 0.8 kW, a chamber pressure of $2.5 \times 10^{-3}$ mbar, a gas flow rate of 300 sccm of argon, and a substrate temperature of 25°C. For the 5 nm of TaN deposition, the DC sputtering power was set at 2.5 kW, with 100 sccm of argon and 180 sccm of $N_2$, and two passes were made at a speed of 56.35 mm/s. The 5 nm TiN deposition was performed at 2.0 kW, with 300 sccm of argon, and 10 passes were conducted at 40 mm/s. Both depositions were carried out at room temperature.

*Surface characterization*

The surface morphologies of Ta films grown on Si surfaces with different seed layers were obtained using the Bruker Dimension Fast Scan AFM with tapping mode tips (FASTSCAN-A). The AFM software NanoScope Analysis (version 1.4) was utilized to analyze the AFM images, enabling the assessment of surface roughness and pinhole depth on the Ta film surface. The impact of changes in the crystal structure and phase formation of the Ta film before and after oxygen plasma exposure was evaluated using XRD (D8 Advance Bruker). Additionally, the surface chemical composition of the Ta thin film and the alterations in the surface oxide composition post-exposure to oxygen plasma treatment were characterized through XPS. The spectra were obtained using a VG ESCA Lab-220i XL XPS system equipped with a monochromatic Al Kα X-ray source with a photon energy of 1486.6 eV at 15 kV. High-resolution spectra were collected at a pass energy of 20 eV, with 0.1 eV steps, at a 45° takeoff angle. The binding energies were corrected to the Carbon 1$s$ energy of 285.0 eV. The collected XPS high-resolution spectra were analyzed using XPS Peak Fit 4.1 software with the Voigt function. Shirley-type backgrounds were subtracted from each spectrum to eliminate most of the extrinsic loss structure.

*TEM sample preparation and data collection*

The lamellae samples were prepared using the FEI Helios 450 dual-beam focused ion beam. The sample was initially coated with 20 nm of carbon, followed by 100 nm of gold to safeguard the top surface during the ion milling process. The samples were then milled to a thickness of less than 70 nm. The primary $Ga^+$ ion beam operated at 30 kV, and 2 kV $Ga^+$ ions were used to clean the samples and remove surface damage or amorphization in the areas of interest. For transmission electron microscopy (TEM) imaging, we utilized an 80-300 kV FEI Titan at 200 kV.

*Critical temperature measurement*

To determine the transition temperature of the Ta film with different seed layers, we measured the resistance across patterned Ta films inside an ICE Oxford cryogenic fridge. We used Python code to record the fridge temperature over time as we slowly heated and cooled the sample around the critical temperature. Simultaneously, we recorded the resistance vs. time values by automatically fitting the current-voltage sweeps measured with a source measure unit (Keithley 2450) controlled by Python-based software. This step was necessary to reduce errors caused by environmental voltage offsets. By synchronizing both records, we obtain the resistance as a function of temperature.

## 3. Results and discussion

The crystal structure and phases of the αTa thin films grown at room temperature on silicon substrates using amorphous and crystalline seed layers were analyzed using XRD. **Figure** 1 shows the XRD patterns of Ta thin films with different seed layers. The spectra confirm the formation of the α-phase Ta film, as evidenced by the (110) phase orientation peak at 38.31° and the second order (220) peak at 82.02°. Notably, no other peaks were observed apart from the (110) and (220) peaks, particularly those associated with weak, different phases of α-Ta or β-Ta peaks previously reported on silicon surfaces.[12] This indicates the successful deposition of a pure α-phase Ta film with a preferred (110) and (220) orientation on Si (100) substrates. This suggests that all seed layers effectively promoted the growth of the α-phase Ta by reducing lattice mismatch with the substrate. In contrast, recent studies by Wu et al.[23] and Singer et al.[12] reported that Ta thin films grown with metal nitride seed layers at room temperature exhibited a majority α-phase but with weak Ta (100) and Ta (200) peak orientations. However, our study found no evidence of additional orientations, including the Ta (100) peak, for the seed layers used.

The inset in **Figure 1** highlights a zoomed-in view of the (110) peak, showing increased intensity and a sharper profile for films grown on crystalline seed layers compared to those grown on amorphous seed layers. The XRD peak shifts to a lower angle, and the FWHM decreases from 1.01° to 0.51°. This indicates periodic crystal arrangement, reduced defects, and highly dense packaging of the films, which minimizes surface defects. While these FWHM values are slightly higher than those previously reported for α-Ta grown on amorphous seed layers, they are significantly sharper than those reported for Nb seed layers (FWHM = 4.23°). Additionally, the crystallite size was reduced when metallic seed layers were used (<15.5 nm) compared to amorphous metal nitride seed layers (>17.4 nm). This suggests that α-Ta thin films achieve denser and more structurally oriented growth with crystalline metallic seed layers. The enhanced relative intensity and sharper FWHM of the (110) peak further confirm grain size improvements and structural modifications influenced by the choice of seed layer.

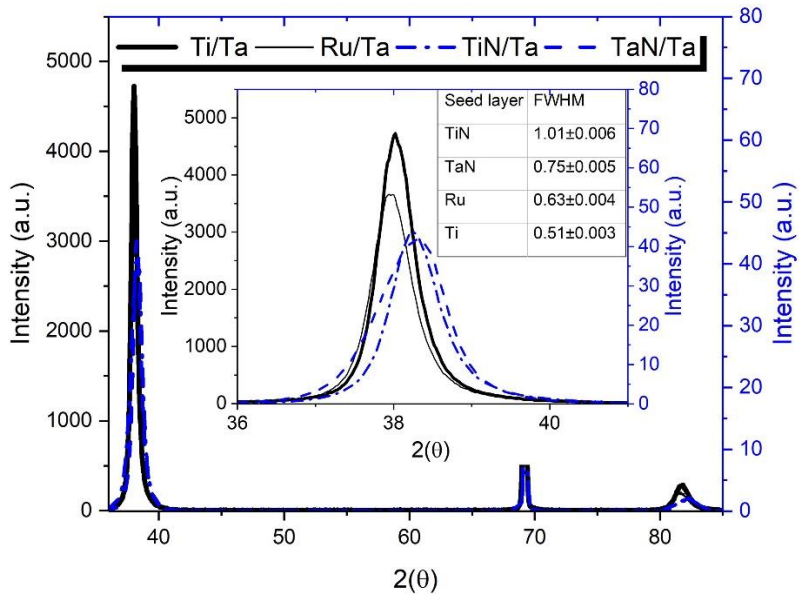

**Figure 1**: The XRD patterns for the room temperature growth of α-Ta on a silicon surface with different seed layers: A) Ti, B) Ru, C) TiN, and D) TaN. The inset shows the zoomed image of a 110 XRD pattern, full-width, and half-maximum of the low-angle peak.

We used tapping mode AFM to study the surface morphology and understand the influence of seed layers on the surface characteristics of the α-Ta film. **Figure 2** presents the AFM images of the α-Ta film grown on different seed layers, revealing elongated crystalline structures with an RMS roughness of ~1.5 nm over a 1×1 μm area for films grown on TiN and TaN seed layers, consistent with previously reported literature. However, distinct surface morphology is observed for metallic crystalline seed layers: the elongated crystalline structures disappear using a Ti seed layer and show RMS roughness of 0.75 nm. In contrast, the Ru seed layer results in densely packed shorter elongated crystalline structures with an RMS roughness of less than 0.80 nm. Thus, we further confirm that α-Ta thin films exhibit denser and more structurally oriented growth when deposited on crystalline metallic seed layers.

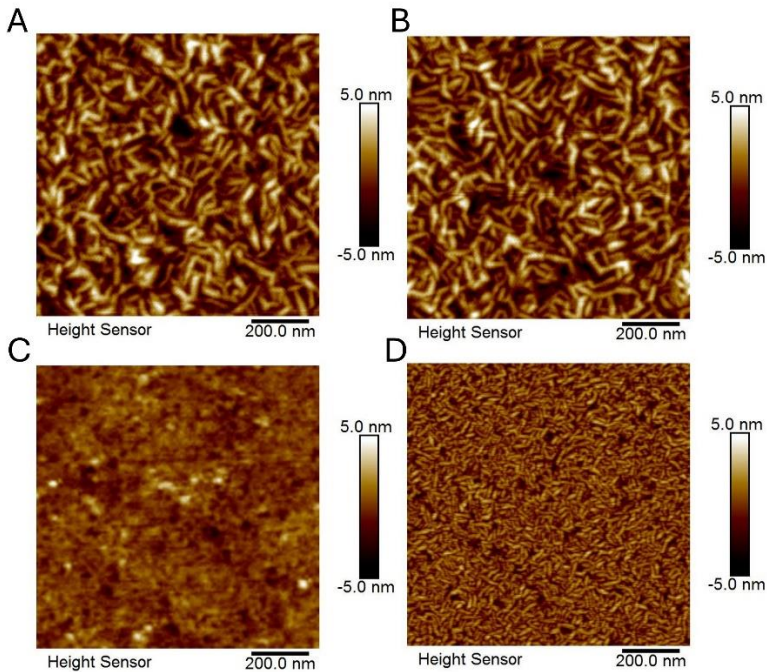

**Figure 2**: AFM image of the room temperature growth of α-Ta on a silicon surface with different seed layers: A) Ti, B) Ru, C) TiN, and D) TaN.

We further analyzed the AFM images to understand the quality of the thin film, and **Figure 3** presents the plot of pinhole depth versus counts for α-Ta thin films grown with different seed layers. The red dotted line represents the Gaussian fitting of the data. The results show that metal nitride seed layers exhibit a pinhole depth of 6.5 ± 1.6 nm. In comparison, the Ti seed layer minimizes the pinhole depth with a low error of 5.3 ± 0.6 nm. Notably, the Ru seed layer achieves a significantly reduced pinhole depth of 2.9 ± 0.7 nm. Thus, the crystalline seed layer demonstrates a lower pinhole depth than the amorphous seed layer, indicating a high-density packing layer of the α-Ta thin film, reducing the surface defect density.

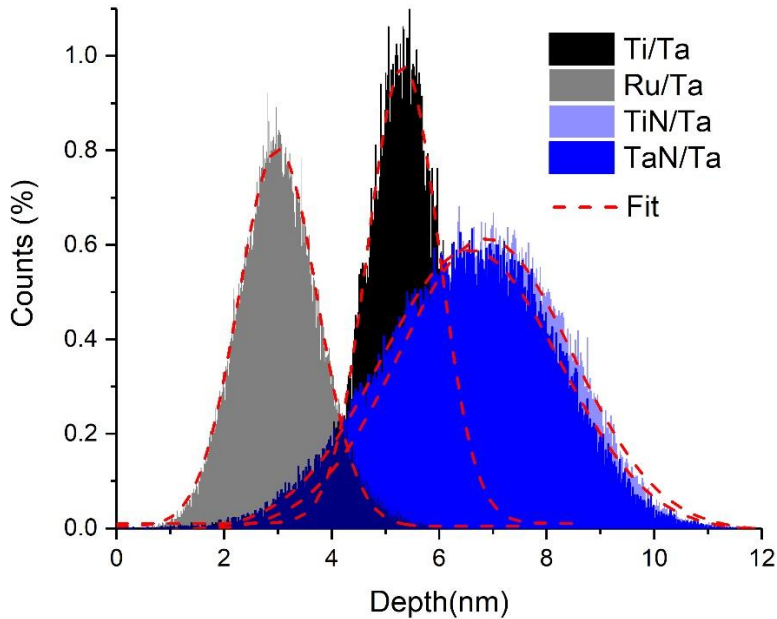

**Figure 3**: Depth profile obtained from the AFM images of the room temperature growth of α-Ta on a silicon surface with different seed layers: A) Ti, B) Ru, C) TiN, and D) TaN.

X-ray photoelectron spectroscopy (XPS) was used to investigate the effect of the seed layer on the surface oxide composition of α-Ta thin films. **Figure 4** shows the Ta 4*f* and O 1*s* spectra of α-Ta films grown with different seed layers while maintaining the same deposition temperature and cleaning process. α-Ta grown on an amorphous seed layer exhibits a higher TaO$_x$ composition (25–29 eV) than the metallic component (20–24 eV), indicating a greater oxide thickness than α-Ta grown on a metallic seed layer. This observation confirms that the atomic packing density is higher in films grown on crystalline seed layers than those on amorphous seed layers. The metal-to-oxide ratio was estimated from the total area of the oxide and metallic components. The Ta surface metal-to-oxide ratio was around 1.3–1.4 for samples with amorphous seed layers, whereas samples with metallic seed layers showed a lower ratio of 0.5–0.74. These results demonstrate that the densely packed Ta structure formed on crystalline seed layers reduces surface oxide thickness compared to the loosely packed structure formed on amorphous seed layers.

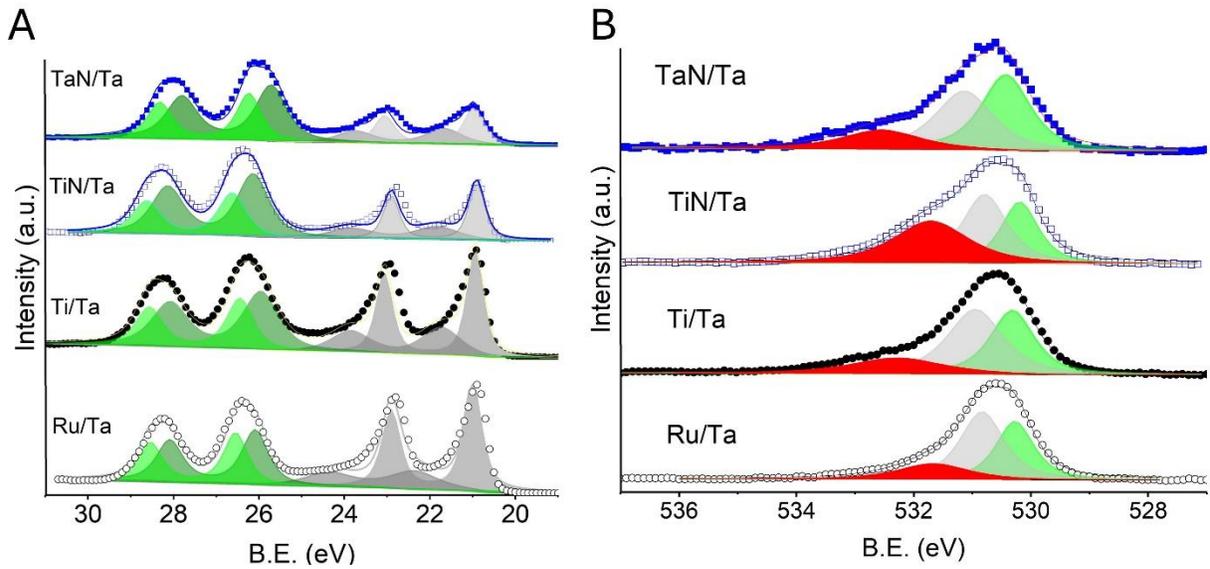

**Figure 4:** The X-ray photoelectron spectra of A) Ta 4$f$ and B) O 1$s$ for the Ta thin film surface grown on a silicon surface with different seed layers.

**Figure 5** presents high-resolution cross-sectional transmission electron microscopy images of the α-Ta film deposited on a Si substrate using various seed layers. **Figures 5(A-B)** illustrate different metallic nitride seed layers, which exhibit an amorphous structure during the deposition process.

Since N₂ gas is used as the iridizing source, achieving defect-free crystalline growth in the 5 nm seed layers is challenging. This observation aligns with previous reports, suggesting that the formation of an amorphous interlayer plays a crucial role in influencing the growth mode and structural properties of the α-Ta film.[12,24] Notably, TaN and TiN have been reported to introduce additional interfaces that may act as potential sources of loss channels due to defects present in the amorphous seed layers.[12] This highlights the need to explore alternative seed layer materials for superconducting quantum computing systems. **Figures 5(C-D)** show metallic seed layers that promote the formation of a crystalline interface between the Si substrate and the Ta thin film. This crystalline interface not only facilitates the growth of the α-Ta phase but also minimizes defects and impurities at the interface compared to amorphous metal nitride seed layers. These characteristics are essential for achieving high-quality superconducting films with reduced loss and improved performance in quantum applications.

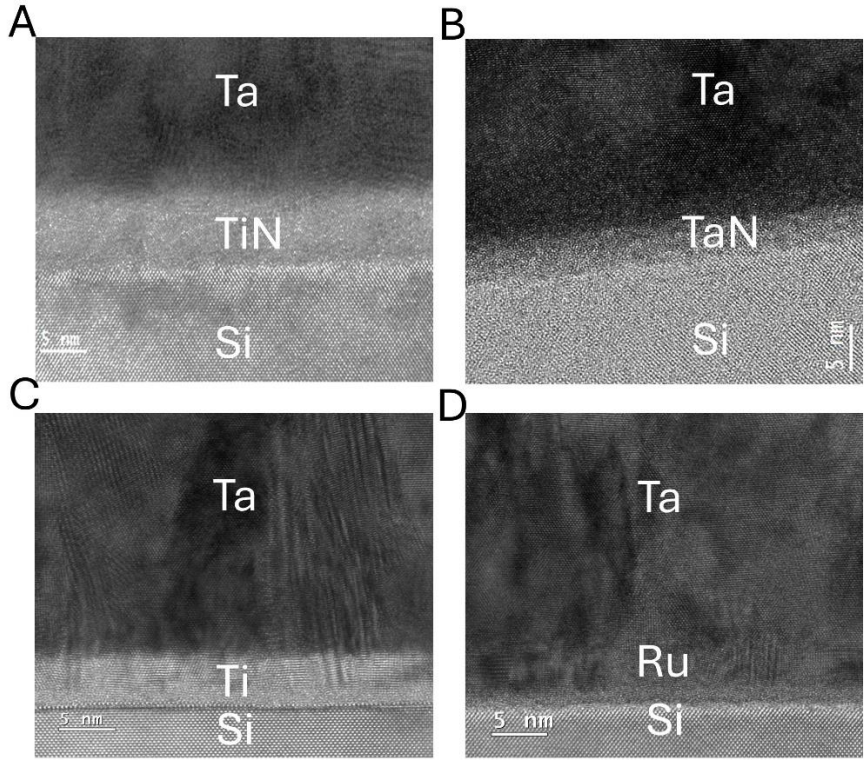

**Figure 5.** Cross-sectional TEM image of the metal-substrate interface for an α-Ta thin film grown on a Si substrate with different seed layers: (A) TiN, (B) TaN, (C) Ti, and (D) Ru.

The electrical properties were measured from patterned α-Ta films grown at room temperature with pre-deposited amorphous and crystalline seed layers. **Figure 6** shows the resistance as a function of temperature. The α-Ta film grown on a pre-deposited amorphous TaN seed layer exhibits a superconducting $T_c$ of 3.91 K, which is slightly lower than the Tc of 3.98 K observed for the TiN seed layer. This result is consistent with previous reports and can be attributed to the lower defect density in TiN compared to TaN. In contrast, α-Ta films grown on metallic seed layers show higher Tc values. The Ti seed layer yields a $T_c$ of 4.15 K, while the Ru seed layer results in a $T_c$ of 4.05 K. These values represent an improvement over previously reported transition temperatures for α-Ta films grown on Si substrates with amorphous seed layers.

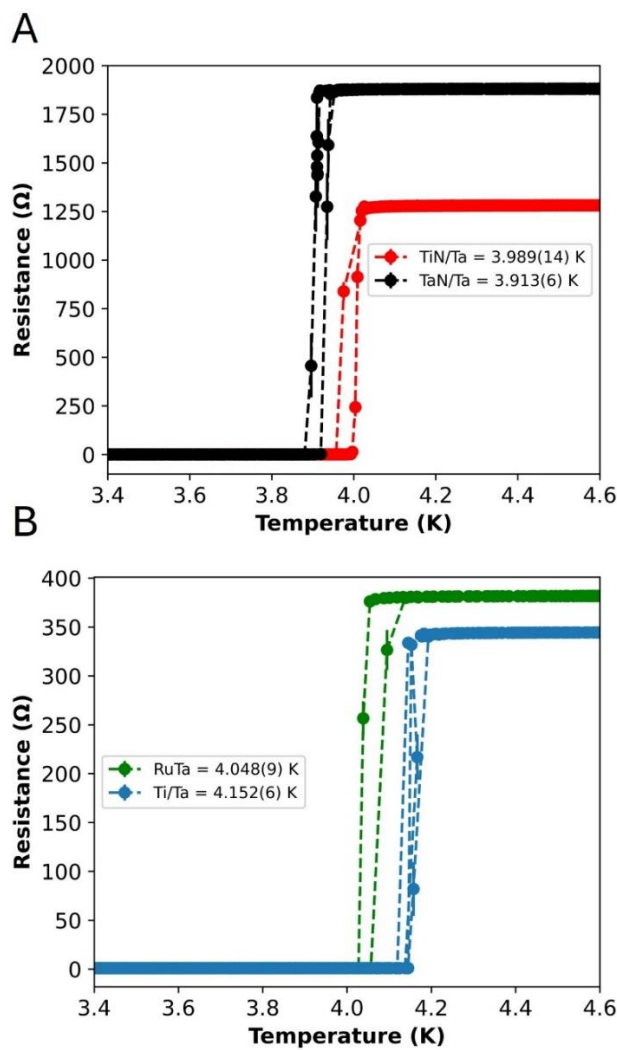

**Figure 6.** Temperature dependence of the electrical transport properties of α-Ta films deposited on a high-resistivity Si substrate with different pre-deposited seed layers: (A) metal nitride and (B) metallic, grown at room temperature.

## 4. Conclusion

The growth and superconducting properties of α-Ta films deposited on high-resistivity Si(100) substrates with crystalline metallic and amorphous buffer layers have been comprehensively studied. XRD, XPS, and AFM analyses reveal that α-Ta films grown with crystalline metallic seed layers exhibit smoother surfaces, higher film quality, and fewer surface defects, as confirmed by lower surface roughness, a thinner oxide layer, a higher metallic component in XPS spectra, and a lower full-width at half-maximum (FWHM) value in XRD peaks. Resistance-temperature measurements indicate that a smoother surface and reduced surface oxide contribute to lower surface resistance, while fewer surface defects correlate with higher superconducting $T_c$ and lower residual resistivity ratio values. These findings provide deeper insights into the relationship between superconductivity and film quality. Our results demonstrate that by carefully controlling the film growth process, high-quality α-Ta films with sharp interfaces can be obtained on Si(100) substrates, achieving both superior material quality and strong superconducting properties, making them highly suitable for integration into large-scale superconducting qubit devices.

## AUTHOR INFORMATION

**Corresponding Author**

karuppannansk@imre.a-star.edu.sg and manas_mukherjee@imre.a-star.edu.sg

**Author Contributions**

The manuscript was written with contributions from all authors. All authors have approved the final version of the manuscript.

**Acknowledgment**

This research is supported by the National Research Foundation, Singapore, and A*STAR under its Quantum Engineering Programme (NRF2021-QEP2-03-P07/W21Qpd0307) and A*STAR SRP (C222517002).

Silicon Wafers: Fabrication, Characterization and Surface Modification. *Mater. Quantum Technol.* **2024**, *4* (2). https://doi.org/10.1088/2633-4356/ad4b8c.